\documentclass[conference]{IEEEtran}
\IEEEoverridecommandlockouts
\usepackage{cite}
\usepackage{amsmath,amssymb,amsfonts}
\usepackage{algorithmic}
\usepackage{graphicx}
\usepackage{textcomp}
\usepackage[table]{xcolor}
\usepackage{booktabs}
\usepackage{comment}
\usepackage[T1]{fontenc}

\usepackage[colorlinks,hidelinks]{hyperref}
\usepackage[capitalize,noabbrev,nameinlink]{cleveref}

\def\BibTeX{{\rm B\kern-.05em{\sc i\kern-.025em b}\kern-.08em
    T\kern-.1667em\lower.7ex\hbox{E}\kern-.125emX}}
\begin{document}

\title{\huge A survey study of success factors in data science projects}

\author{

\IEEEauthorblockN{I\~nigo Martinez}
\IEEEauthorblockA{\textit{Vicomtech Foundation} \\
\textit{Basque Research and Technology Alliance}\\
Donostia-San Sebastián 20009, Spain \\
imartinez@vicomtech.org}
\and
\IEEEauthorblockN{Elisabeth Viles}
\IEEEauthorblockA{
\textit{University of Navarra}\\
\textit{TECNUN School of Engineering} \\
Donostia-San Sebastián 20018, Spain \\
eviles@tecnun.es}
\and
\IEEEauthorblockN{Igor G Olaizola}
\IEEEauthorblockA{\textit{Vicomtech Foundation} \\
\textit{Basque Research and Technology Alliance}\\
Donostia-San Sebastián 20009, Spain \\
iolaizola@vicomtech.org}
}

\IEEEpubid{978-1-6654-3902-2/21/\$31.00 \copyright2021 IEEE}

\makeatletter
\def\ps@IEEEtitlepagestyle{
  \def\@oddfoot{\mycopyrightnotice}
  \def\@evenfoot{}
}
\def\mycopyrightnotice{
  {\footnotesize
  \begin{minipage}{\textwidth}
  \centering
  978-1-6654-3902-2/21/\$31.00 \copyright2021 IEEE \\~\\
  \copyright 2021 IEEE.  Personal use of this material is permitted.  Permission from IEEE must be obtained for all other uses, in any \\ current or future media, including reprinting/republishing this material for advertising or promotional purposes, creating new \\ collective works, for resale or redistribution to servers or lists, or reuse of any copyrighted component of this work in other works.
  \end{minipage}
  }
}

\maketitle

\begin{abstract}
In recent years, the data science community has pursued excellence and made significant research efforts to develop advanced analytics, focusing on solving technical problems at the expense of organizational and socio-technical challenges. 
According to previous surveys on the state of data science project management, there is a significant gap between technical and organizational processes. 
In this article we present new empirical data from a survey to 237 data science professionals on the use of project management methodologies for data science. We provide additional profiling of the survey respondents' roles 
and their priorities when executing data science projects.
Based on this survey study, the main findings are: 
(1) Agile data science lifecycle is the most widely used framework, but only 25\% of the survey participants state to follow a data science project methodology. 
(2) The most important success factors are precisely describing stakeholders' needs, communicating the results to end-users, and team collaboration and coordination. 
(3) Professionals who adhere to a project methodology place greater emphasis on the project's potential risks and pitfalls, version control, the deployment pipeline to production, and data security and privacy.
\end{abstract}

\begin{IEEEkeywords}
data science, survey, project management, factor analysis, success factors
\end{IEEEkeywords}

\section{Introduction}\label{sec:introduction}

In recent years, the data science (DS) community has pursued excellence and made significant research efforts to develop advanced analytics, focusing on solving technical problems at the expense of organizational and socio-technical challenges. Developing and successfully implementing a data science project to business processes pose significant challenges to companies. According to VentureBeat \cite{venturebeat_ds_project}, 87\% of data science projects never make it into production, and a NewVantage survey \cite{new_vantage_summary} found that adoption of big data and artificial intelligence (AI) initiatives remain a significant challenge for 77\% of businesses. In fact, 70\% of companies report minimal or no impact from AI \cite{ransbotham_2019}. Gartner \cite{gartner_advanced_analytics} also reported that 80\% of analytics insights will fail to deliver business outcomes by 2022, and 80\% of data science projects will ``remain alchemy, run by wizards''. These low success rates in data science projects are indeed noteworthy given the competitive advantage that such advanced techniques provide to researchers and practitioners. 

Leveraging data science within a business organizational context involves organizational and socio-technical challenges beyond the analytical ones: a lack of vision, strategy and clear objectives, a biased emphasis on technical issues, a lack of reproducibility, and role ambiguity are among these challenges, which lead to a low level of maturity in data science projects, that are managed in an ad-hoc fashion. The lack of a defined process methodology to manage data science projects is one of the primary reasons for these challenges. \cite{saltz_new_processes}. 

Several surveys \cite{piatetsky_crispdm, Corinium, saltz_crispdm, saltz_2021} have been conducted to collect empirical data on the state of data science project management in both industry and research. According to the findings of these surveys, there is a significant gap between technical and organizational processes when applied to data science projects. 

In this article, we present new empirical data from a survey to 237 data science professionals on the use of project management methodologies for data science. Furthermore, we provide additional profiling of the survey respondents' roles (data scientist, data engineer, business analyst, etc.) and their priorities when executing data science projects. 

The rest of the article is structured as follows:  
Background context is presented in \cref{sec:background}, along with organizational and socio-technical challenges that arise when executing a data science project. 
\cref{sec:methods} describes the methodology for the survey.
\cref{sec:results} presents and discusses the results obtained from the survey and finally, \cref{sec:conclusions} describes the main conclusions of the paper.

\section{Background Context}\label{sec:background}

Data science is a multidisciplinary field that lies between computer science, mathematics and statistics, and comprises the use of scientific methods and techniques, to extract knowledge and value from large amounts of structured and/or unstructured data. 
After several decades of progress, data science and machine learning technology are called to become a significant source of value for a wide range of businesses \cite{state_of_ai, Mondal2020}. 
However, these recent technological achievements do not correspond to their application to real-world data science projects. Data scientists are struggling to deploy their models into business processes. Evidence suggests that the gap is widening between organizations successfully gaining value from data science and those struggling to do so \cite{mitsloan}. Significant challenges remain unsolved, and many artificial intelligence initiatives fail.  

\begin{table*}[hbt!]
	\centering
	\resizebox{1\textwidth}{!}{
	\begin{tabular}{@{}l|l|l@{}}
		\toprule
		\textbf{Team Management}                & \textbf{Project Management}                   & \textbf{Data \& Information Management} \\ 
		\midrule
		Poor coordination                       & Low level of process maturity                 & Lack of reproducibility                 \\
		Collaboration issues across teams       & Uncertain business objectives                 & Retaining and accumulation of knowledge \\
		Lack of transparent communication       & Setting adequate expectations                 & Low data quality for ML                 \\
		Inefficient governance models           & Hard to establish realistic project timelines & Lack of quality assurance checks        \\
		Lack of people with analytics skills    & Biased emphasis on technical issues           & No validation data                      \\
		Rely not only on leading data scientist & Delivering the wrong thing                  & Data security and privacy               \\
		Build multidisciplinary teams           & Project not used by business                  & Investment in IT infrastructure         \\ 
		\bottomrule
	\end{tabular}
	}
	\vspace{1em}
	\caption{Data science projects main challenges}
	\vspace{-2em}
	\label{tab:challenges}
\end{table*}

\newpage
\cref{tab:challenges} summarizes the main challenges faced by data science professionals when executing real data science projects.
Even though these issues do exist in real-world data science projects, the community has not been overly concerned about them \cite{saltz_new_processes, saltz_sociotechnical_challenges}, and there is a wide consensus that not enough has been written about solutions to tackle these problems \cite{MARTINEZ2021100183}. 

Some authors and companies have proposed methodologies for managing data science projects, and have come up with new tools and processes to address the cited issues. The most popular frameworks in this regard are: CRISP-DM \cite{wirth2000crisp}, Microsoft TDSP \cite{microsoft}, Agile DS Lifecycle \cite{russell2017agile}, Domino DS Lifecycle \cite{hotz_dominolab}, IBM FMDS \cite{rollings_foundational_ibm}, RAMSYS \cite{moyle_ramsys}, and MIDST \cite{crowston_midst}. An extensive critical review over 19 data science methodologies was presented in \cite{MARTINEZ2021100183}.

However, while the proposed solutions are leading the way in addressing these organizational and socio-technical challenges, the reality is that data science projects are not taking advantage of such methodologies and frameworks. According to a survey \cite{saltz_exploring_pm} carried out in 2018 to professionals from both industry and non-profit organizations, 82\% of the respondents did not follow an explicit process methodology for developing data science projects, and equally important, 85\% of the respondents considered that using an improved and more consistent process would result in more effective data science projects. Furthermore, a survey conducted to nearly 800 data and analytics leaders by Corinium Intelligence \cite{Corinium} found that only 48\% of data science organizations have established standardized processes.

Considering a survey from KDnuggets in 2014 \cite{piatetsky_crispdm}, the main methodology used by 43\% of respondents was CRISP-DM. This methodology has been consistently the most commonly used for analytics, data mining, and data science projects, for each KDnuggets poll starting in 2002 up through the most recent 2014 poll \cite{saltz_crispdm}. 
Recently, Saltz \cite{saltz_2021} conducted a survey to 109 respondents to gather information about what project management framework teams used to help them execute data science projects. CRISP-DM is used by nearly half of all respondents. Scrum, Kanban, and customized processes followed. 

According to the results of these surveys, the percentage of people using CRISP-DM has not changed significantly over the last 20 years, indicating that technology has advanced much faster than organizational processes for handling data science projects.


\newpage
\section{Methods}\label{sec:methods}

To better understand the use of project management methodologies and the priorities of data science professionals during project execution, a cross-sectional quantitative survey was conducted. This survey focused on understanding the professional profile of the respondents ($Q_{1}-Q_{5}$), the type of projects they carry out ($Q_{6}-Q_{8}$), the most important factors to successfully develop a DS project $Q_{9}$ and whether they use any project methodology $Q_{10}-Q_{11}$ (\cref{tab:questions}). In total 237 professionals from industry and research organizations were surveyed during September 2021. Participants accessed the survey via the LinkedIn professional network.

\begin{table}[!hbt]
\centering
\resizebox{1\linewidth}{!}{%
\begin{tabular}{r|l|c}
\toprule & \textbf{Question} & \textbf{Fig.} \\ \midrule
$Q_1$ & In what country do you work? &  \ref{fig:plot_country}\\
$Q_2$ & Which title best describes your role? &  \ref{fig:plot_roles}\\
$Q_3$ & Which is your job experience level? &  \ref{fig:plot_questions}\\
$Q_4$ & Who are the stakeholders of your work projects? &  \ref{fig:plot_questions} \\
$Q_5$ & What is your gender? &  \ref{fig:plot_questions} \\
$Q_6$ & Average project duration (<3, <6, >6 months) &  \ref{fig:plot_questions} \\
$Q_7$ & Do you usually work in group or individually? &  \ref{fig:plot_questions} \\
$Q_8$ & Magnitude of the project (local/national/international) &  \ref{fig:plot_questions} \\
$Q_9$ & \begin{tabular}[t]{@{}l@{}}How relevant are these factors for the \\ development of data science projects?\end{tabular}  &  \ref{fig:plot_aspects}\\
$Q_{10}$ & \begin{tabular}[t]{@{}l@{}}Do you usually follow some data \\ science project methodology?\end{tabular} &  \ref{fig:plot_questions}\\
$Q_{11}$ & \begin{tabular}[t]{@{}l@{}}Select the methodologies you know about \\ and/or have used for your data science projects\end{tabular}  &  \ref{fig:plot_methodology_use}\\ \bottomrule
\end{tabular}%
}
\vspace{1em}
\caption{Survey questions}
\vspace{-2em}
\label{tab:questions}
\end{table}


\section{Results}\label{sec:results}

The results of the survey are presented and discussed in this section.
First, the respondents' professional profiles are examined.
The use of data science methodologies is then investigated, and finally, the relevance scores assigned to fifteen success factors are presented. 

\subsection{Professional profiles}\label{subsec:profiles}

The survey included professionals from 35 different countries, the majority of whom work in the United States of America or Spain. 
Among the survey respondents, 27\% were data scientists, 14\% data engineers, 12\% business analysts, 12\% data analysts, 11\% machine learning engineers and 6\% software developers. \cref{fig:plot_country,fig:plot_roles} show a detailed view of work-country and professional-roles respectively. 

\begin{figure}[!hbt]
    \centering
	\includegraphics[width=1\linewidth]{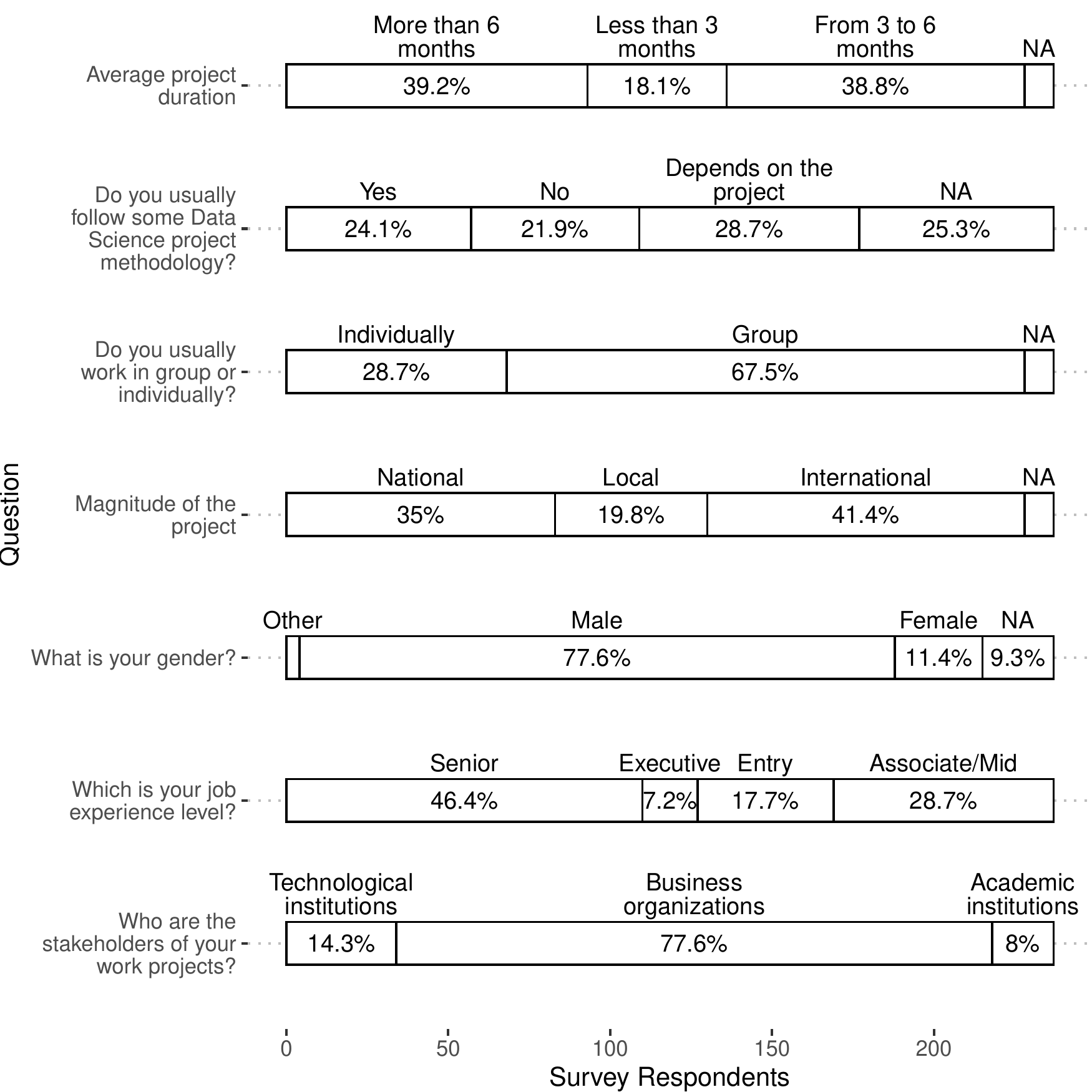}
	\caption{Survey results to additional questions $Q_{3}-Q_{8}, Q_{10}$. NA means \textit{"Not Answer"}.}
	\label{fig:plot_questions}
\end{figure}

The vast majority of respondents (77\%) identified as male. The level of job experience among the professionals polled was distributed as follows: 46\% senior, 28\% associate, 18\% entry-level, and 7\% executive. Hence, more than half of the surveyed professionals have many years of experience under the belt.

With regard to project characteristics, nearly 40\% of the respondents say a project usually takes more than 6 months to complete, while another 40\% says it takes between 3 to 6 months. This may suggest that long projects are the norm in the field of data science. In terms of projects magnitude, 42\% of the professionals work on international projects, 35\% on national and 20\% on local projects. This implies that the majority of data science projects are large and involve a large number of stakeholders, who in the case of this survey are mostly business organizations (77\%). With respect to teamwork, 68\% of the professionals polled say they usually work in groups, with 29\% working individually. This shows the importance of coordination, transparent communication, and collaboration among individuals and teams for the success of DS projects. See \cref{fig:plot_questions} for a detailed view on these matters.

\subsection{Data science methodologies}\label{subsec:ds_methodologies}

\begin{figure}[!hbt]
    \centering
	\includegraphics[width=1\linewidth]{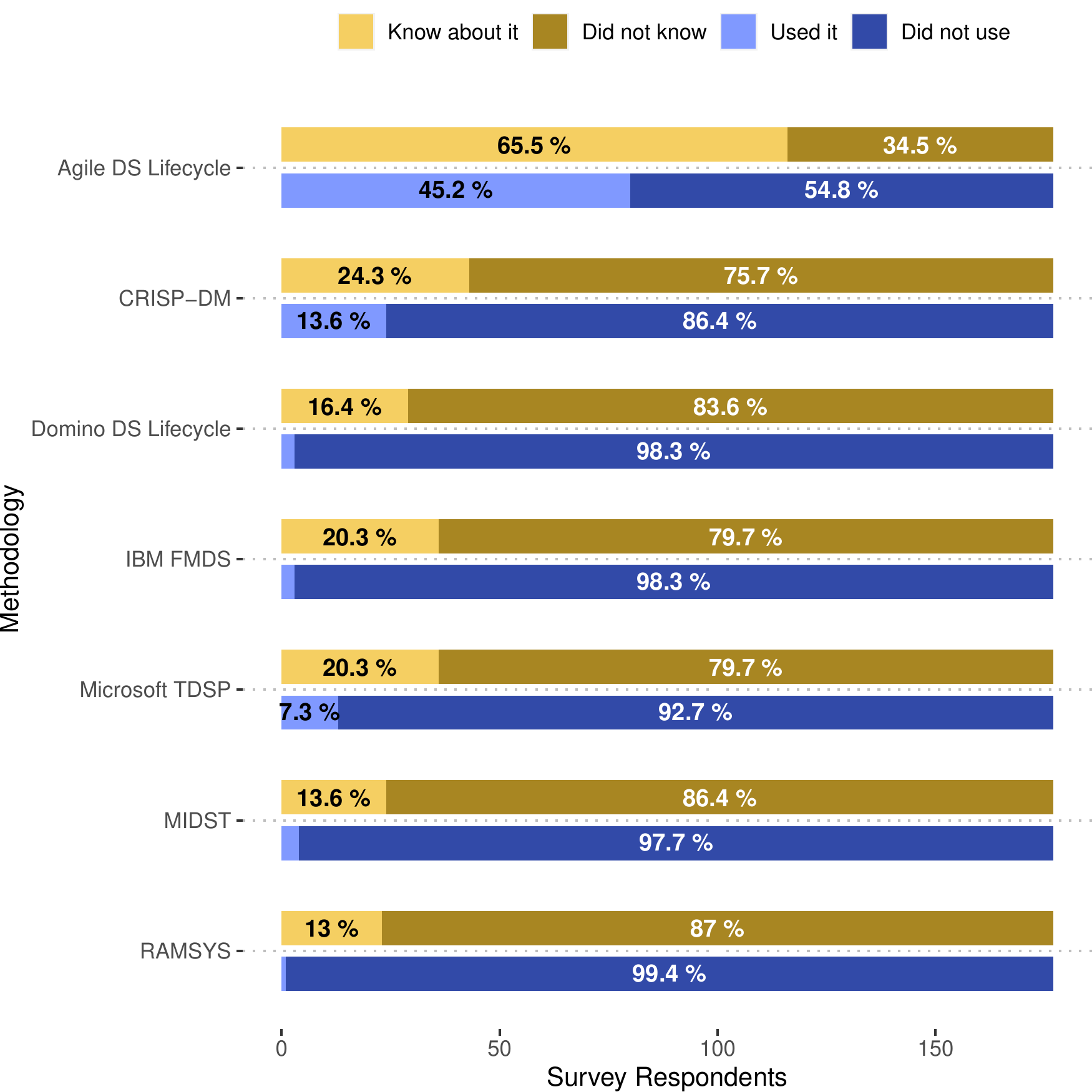}
	\caption{Survey results to the use of methodologies for data science projects (question $Q_{11}$)}
	\label{fig:plot_methodology_use}
\end{figure}

With regard to data science methodologies, \textit{Agile DS Lifecycle} is the most widely used framework, followed by \textit{CRISP-DM} and \textit{Microsoft TDSP} (\cref{fig:plot_methodology_use}). Few people use their in-house methodology. However, it is remarkable to see that only 25\% of the participants are aware of \textit{CRISP-DM}, whereas 65\% of them know about \textit{Agile DS Lifecycle}.

The fact that 75\% of the respondents are unaware of CRISP-DM may indicate a different trend in the use of this methodology when compared with data collected in previous surveys \cite{piatetsky_crispdm, saltz_2021}, with around 45\% use of CRISP-DM. Other factors, such as the number of respondents and the distribution of their professional profiles, may have skewed these findings. An extensive survey could help to reduce this uncertainty.

Furthermore, only 25\% of the survey participants state they use some kind of data science project methodology, and another 29\% say it depends on the project. Moreover, the number of \textit{NA (not answer)} responses may suggest that the proportion of people using a DS methodology is less than 25\%. In any case, this low percentage demonstrates the lack of a defined process methodology for managing data science projects, which has been recognized as one of the primary causes for current failures and management challenges.

\subsection{Success factors}\label{subsec:success}

\begin{figure*}[!hbt]
    \centering
	\includegraphics[width=1\linewidth]{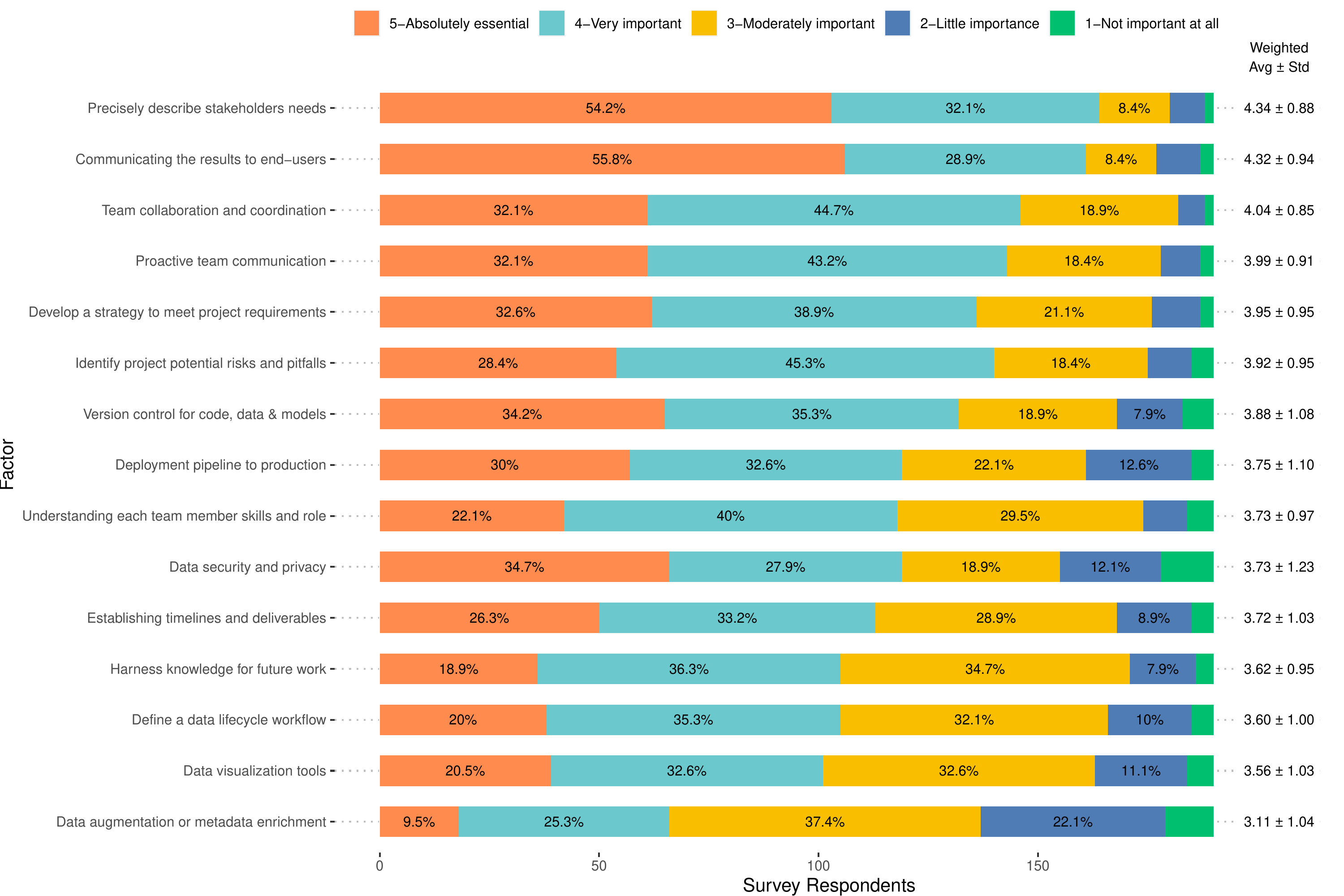}
	\caption{Survey results to \textit{"how relevant are these factors for the development of data science projects?"} question $Q_{9}$. The scoring weights range from 1 (not important at all) to 5 (absolutely essential). The weighted average and standard deviation are calculated based on the response count and the weight for the answer choice.}
	\label{fig:plot_aspects}
\end{figure*}

Question $Q_{9}$ requested survey participants to evaluate the importance of fifteen factors for the development and success of data science projects. These factors are thought to contribute to the project's success in meeting its objectives, and were selected based on an analysis of current challenges in data science projects \cite{MARTINEZ2021100183}.
The scoring system followed a 5 point Likert scale, from 1 (not important at all) to 5 (absolutely essential). \cref{fig:plot_aspects} shows the distribution of the scores, along with the weighted average and standard deviation for each factor. The factors are listed in descending order (from top to bottom in the figure) based on their weighted average score. The top three most important success factors are to 1) \textbf{precisely describe stakeholders needs}, 2) \textbf{communicate the results to end-users}, and 3) \textbf{team collaboration and coordination}. The less important ones, on the other hand, are \textit{data augmentation}, \textit{data visualization tools} and defining \textit{data life-cycle workflow}. It is also worth noting the factors with the highest standard deviations: \textit{deployment pipeline to production} and \textit{data security and privacy} had more variance in their scores, which could indicate that professionals have differing views on the importance of these factors.

\begin{figure}[!hbt]
    \centering
	\includegraphics[width=\linewidth]{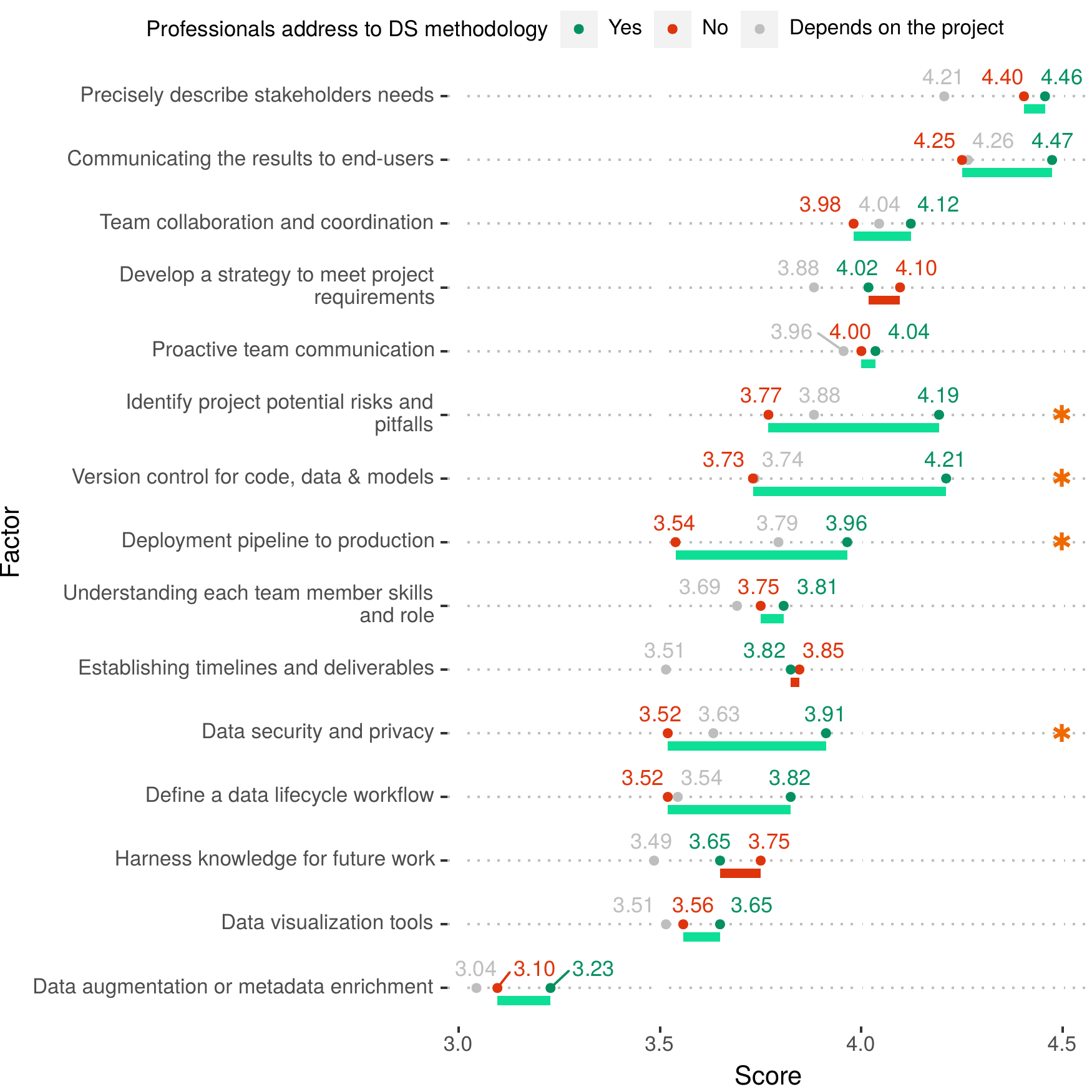}
	\caption{Comparison of weighted average scores between people who follow (and do not) follow any data science project methodology. The orange asterisk on the right side indicates a significant difference between the \textit{Yes} and \textit{No} cases, based on Mann-Whitney U-test.}
	\label{fig:plot_improvement}
\end{figure}

Furthermore, an additional analysis of these scores was performed to determine whether there were any significant differences between professionals who use a project methodology and those who do not. Using the responses from question $Q_{10}$ (\textit{Do you usually follow a data science project methodology?}), the weighted average scores were computed for each choice (\textit{Yes / No / Depends on the project}), which are shown in \cref{fig:plot_improvement}. The orange asterisk on the right side indicates a significant difference between the \textit{Yes} and \textit{No} cases, based on Mann-Whitney U-test. Statistical tests are performed using R v3.6.3 \cite{R}. Professionals who follow a project methodology give more importance to a) the project's potential risks and pitfalls, b) version control, c) the deployment pipeline to production, and d) data security and privacy.

\subsection{Common factor analysis}\label{subsec:factor}

Using the results from question $Q_9$, common factor analysis was conducted to see if there were are any latent variables in the data, i.e., groups of factors that were answered similarly based on the survey participants' scores.

\begin{figure}[!hbt]
    \centering
	\includegraphics[width=1\linewidth]{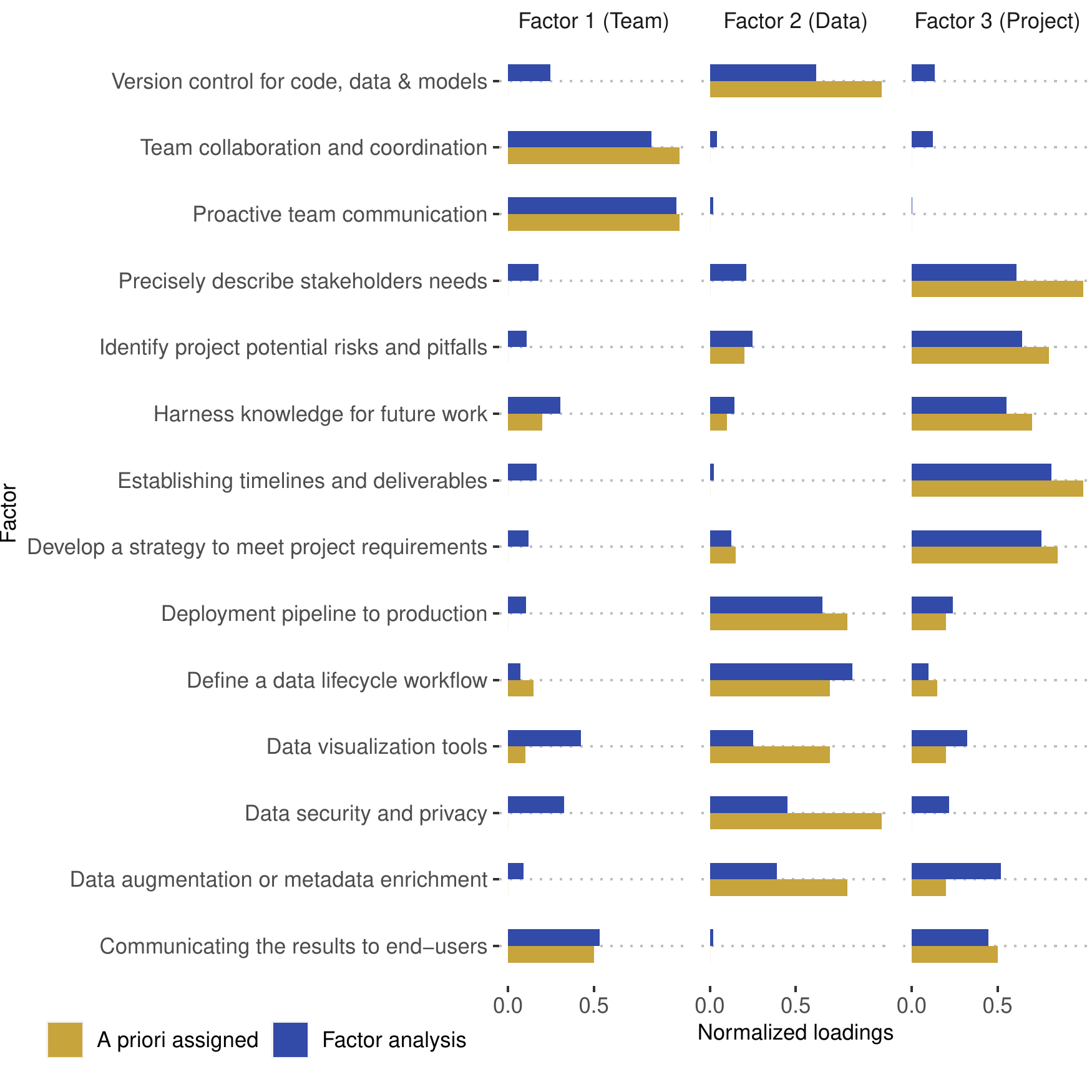}
	\caption{Factor analysis loadings. Comparison with preassigned weights prior to the survey.}
	\label{fig:plot_factor_analysis}
\end{figure}

To begin, Bartlett’s test of sphericity and the Kaiser-Meyer-Olkin measure of sampling adequacy were used to check whether any meaningful latent factors could be found within the data. The parallel analysis method was then used to determine the number of factors, which resulted in 3 factors. The results of the common factor analysis are shown in \cref{fig:plot_factor_analysis} (in blue color). Prior to the survey's public release, each of the fifteen factors was weighted and classified based on three distinct and specific latent dimensions (in yellow). These latent variables were selected based on a previous paper by the authors \cite{MARTINEZ2021100183}, which proposes the principles of an integral methodology for data science that should include three foundation stones: team, data and project management. More information on the proposed principles for integral methodologies can be found in section 5 (discussion) of \cite{MARTINEZ2021100183}. Comparing the results from the common factor analysis with the authors' a priori determined weights, it can be observed that the first factor is related to the \textbf{team management} facet, the second to the \textbf{data management} aspect and the third to the \textbf{project management}. Hence, based on the relevance scores provided by the survey participants and the conducted factor analysis, it can be concluded that there are three latent variables and, in addition, these coincide with the a priori set values based on the proposed framework \cite{MARTINEZ2021100183}. 

\section{Conclusions}\label{sec:conclusions}

This article presents new empirical data from a survey to 237 data science professionals on the use of project management methodologies for data science. Several aspects of survey respondents' work as data science professionals were evaluated, including their experience level and role (data scientist, data engineer, business analyst, etc.), projects characteristics (magnitude, duration) and their priorities when executing data science projects. Regarding the limitations of the presented study, note there may be a bias due to the survey's non-probabilistic design.

According to this survey study, the most widely used framework is \textit{Agile DS lifecycle}, which has surpassed the long-standing CRISP-DM methodology. However, only 25\% of the survey participants state to follow a data science project methodology. This low percentage shows the lack of a defined process methodology for managing data science projects, which has been identified as one of the primary causes for current failures and management challenges. 
With regard to the study of the success factors, based on the scores gathered, the most important factors are (1) precisely describing stakeholders' needs, (2) communicating the results to end-users, and (3) team collaboration and coordination. 
Furthermore, a statistical analysis of the scores revealed that professionals who adhere to a project methodology place a greater emphasis on the project's potential risks and pitfalls, version control, the deployment pipeline to production, and data security and privacy.
The obtained scores were subjected to a common factor analysis, which revealed three latent factors. Comparing the results from the common factor analysis with a priori determined weights, the first factor was related to the team management facet, the second to the data management aspect and the third to the project management. 

Therefore, this survey reinforces the notion that the three most important factors for successful DS projects are project development, teamwork, and data management. The latter is also most valued by those professionals who use DS methodologies. 
However, the methodologies that attempt to address these three factors and were best valued according to the principles of an integral methodology in \cite{MARTINEZ2021100183} are hardly known and used. 
There is still a need for greater dissemination of the methodologies that reinforce these factors or for improving and incorporating such factors into the most widely used methodologies. 


\bibliographystyle{IEEEtran}
\bibliography{IEEEabrv,references}

\newpage
\appendices
\section{Survey additional results}

\setcounter{figure}{0}
\renewcommand\thefigure{\Alph{section}.\arabic{figure}}

\begin{figure}[!hbt]
    \centering
	\includegraphics[width=1\linewidth]{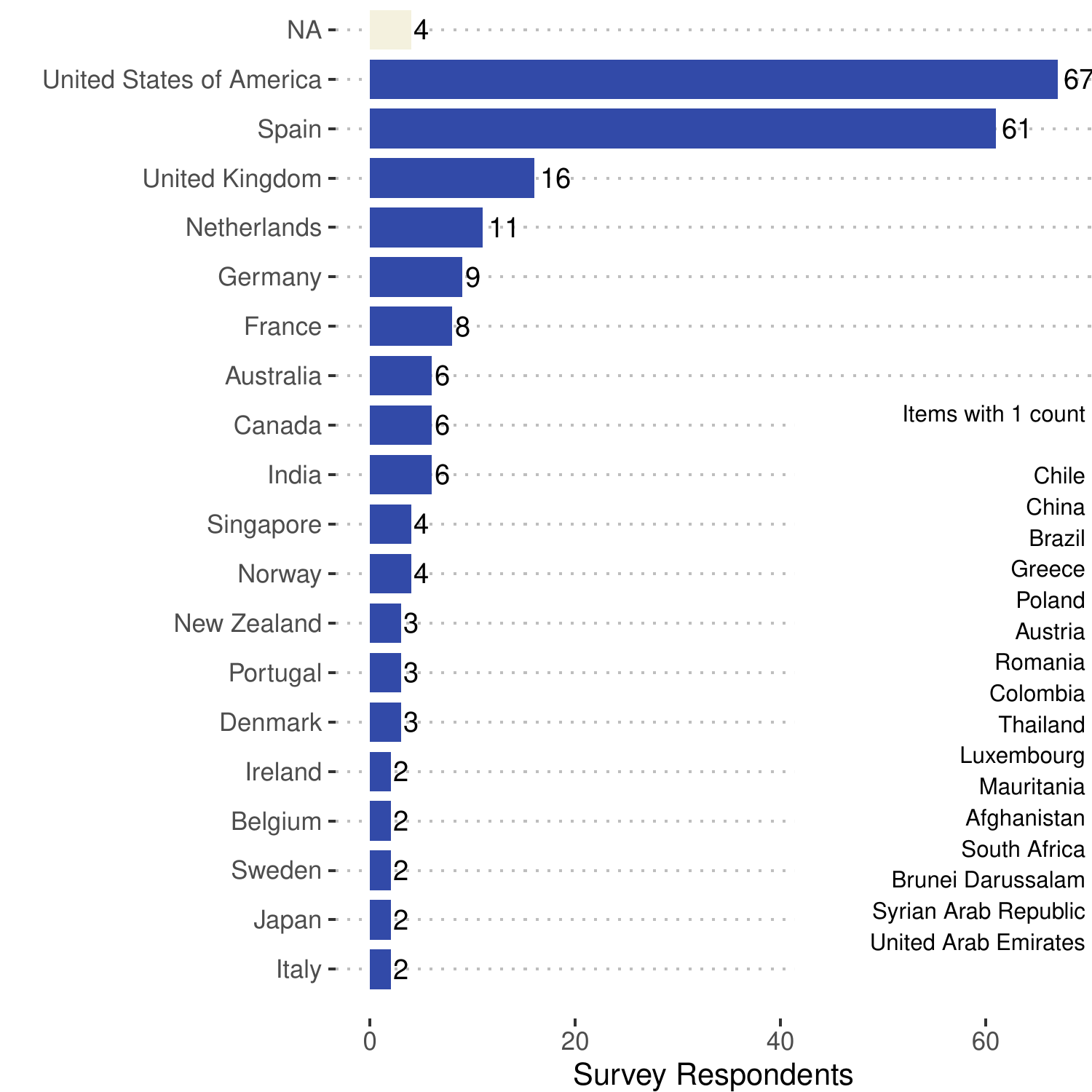}
	\caption{Survey results to \textit{"in what country do you work?"} question $Q_1$}
	\label{fig:plot_country}
\end{figure}

\begin{figure}[!hbt]
    \centering
	\includegraphics[width=1\linewidth]{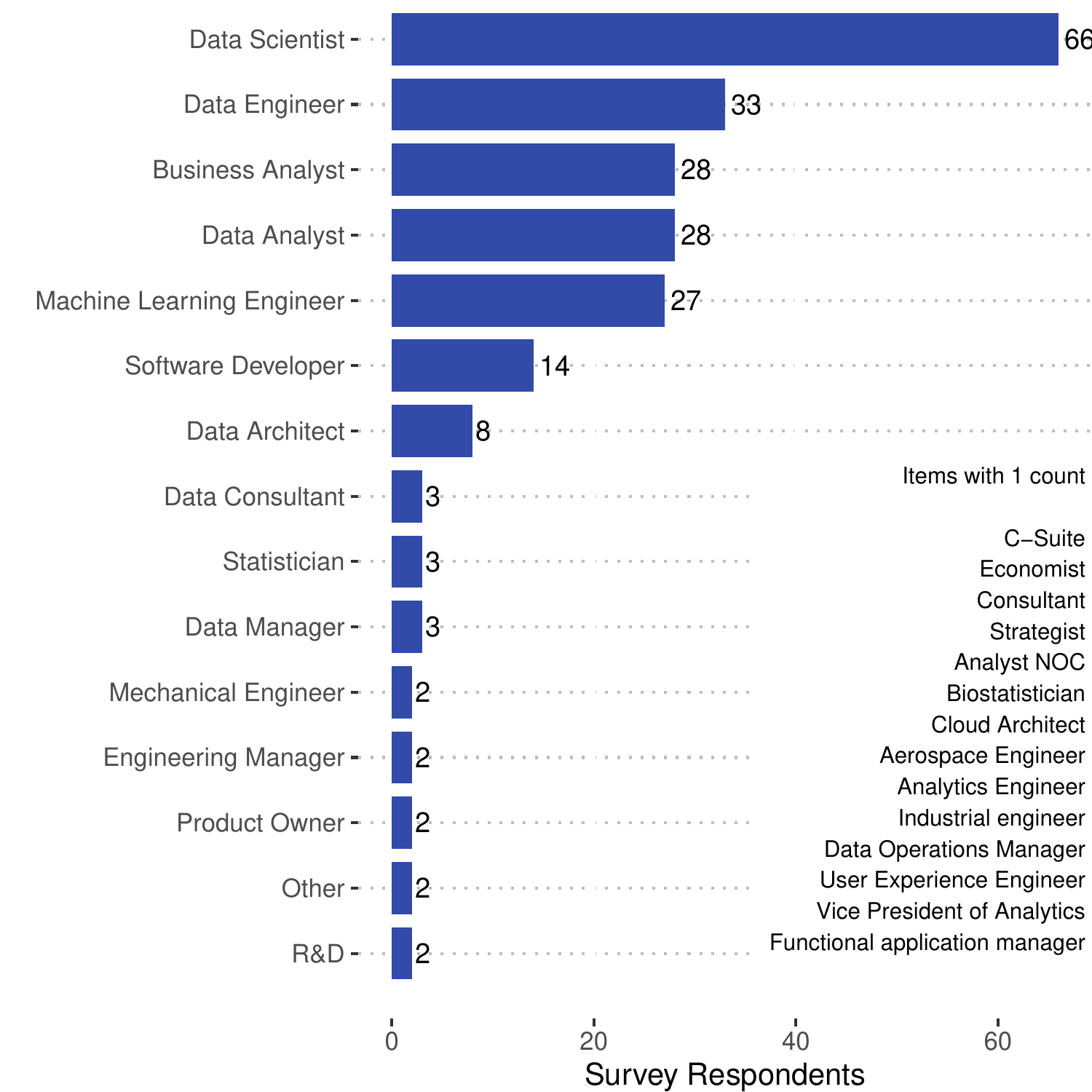}
	\caption{Survey results to \textit{"which title best describes your role?"} question $Q_2$}
	\label{fig:plot_roles}
\end{figure}

\end{document}